\documentclass[11pt]{article}

\usepackage[final]{acl}

\usepackage{times}
\usepackage{latexsym}

\usepackage[T1]{fontenc}

\usepackage[utf8]{inputenc}

\usepackage{microtype}

\usepackage{inconsolata}

\usepackage{xcolor,colortbl}
\usepackage{graphicx}
\usepackage{enumitem}
\usepackage{tabularx}
\usepackage{tabularray}
\usepackage{booktabs}
\usepackage{tcolorbox}
\usepackage{siunitx}
\usepackage{amsmath}
\usepackage{amssymb}
\usepackage{float}
\usepackage{caption}
\usepackage{subcaption}
\usepackage{hyperref}

\title{When RAG Doesn’t Help: LLMs for\\Plausible Material Prediction of Vehicle Components}

\author{Frederik Wagner \and Annerose Eichel \and Sabine Schulte im Walde  \\
         University of Stuttgart, Institute for Natural Language Processing, Germany \\ \normalsize \texttt{\{frederik.wagner,annerose.eichel,schulte\}@ims.uni-stuttgart.de} \\}

\begin{document}
\maketitle
\begin{abstract}
In this work, we explore whether LLMs can predict and explain plausible materials for vehicle components such as \textit{brake discs} or \textit{fuel injectors} without extensive fine-tuning. We compare a standard generative LLM baseline, single-pass Retrieval-Augmented Generation (RAG), and an iterative Chain-of-Verification (CoVe) variant across two LLMs, using a domain-filtered Wikipedia corpus that proved helpful in prior work.
As no gold standard exists, we develop and release a reusable web-based annotation tool supporting crucial functions for structured domain-expert judgments, such as highlighting, ratings and ranking. 
Expert agreement is reasonable for plausibility, lower for ratings. LLM-based generation compares favorably to prior work. RAG and CoVe do not improve on it and are partly rated worse by experts. Contrary to prior work, Wikipedia-based retrieval does not help in our setup.
\end{abstract}

\section{Introduction}
LLMs have become part of everyday life, and are increasingly used in professional settings. This includes high-stakes domains with limited available evaluation of their reliability in corresponding real-life deployment 
\cite{merenda-etal-2026-llms,delmas-etal-2025-accelerating,bakos-etal-2025-generating}. 
The vehicle repair domain is one such real-life scenario, which additionally raises critical concerns in terms of physical safety (e.g., improperly executed repairs may impact the reliability of a vehicle). Our work addresses this gap, focusing on the vehicle repair setting with a need for aid in information retrieval.

Consider a vehicle repair shop where an AI assistant is used to support a mechanic, for example, by guiding them through a repair process. To do this, models underlying an AI assistant need to combine world knowledge with domain-specific information. For example, most people would know that \textit{fabric} is not a plausible material for a \textit{brake disc} based on their general understanding of the world. But without domain-specific background knowledge in the automotive domain, a lay person is likely not able to precisely predict the actual materials that a brake disc can be made out of, such as \textit{gray cast iron} and \textit{stainless steel}. In this study, we use the task of \textbf{predicting plausible materials for vehicle components} to evaluate whether and to which extent LLMs acquire 
useful domain-specific 
knowledge. 
Strong performance on this task could make such a model a valuable expert decision-support tool.

We draw on LLMs that are trained on vast amounts of text, thus learning
world knowledge through distributional patterns \citep{sun-etal-2024-head,holtermann-etal-2025-around}. In our baseline experiment, we test whether an \textbf{off-the-shelf LLM}-based approach can predict plausible materials for vehicle components in the vehicle repair domain without extensive pre-training or fine-tuning.
Our results compare favorably to prior work \citep{schlipf:22, eichel-etal-2023-made}, showing that plausible materials for vehicle components may be elicited from an off-the-shelf LLM with a fair degree of confidence.
However, it remains unclear \textit{why} specific material candidates are predicted.
The second part of this work therefore asks for adding an explanation of each predicted material, i.e., a brief \textbf{rationale linking the material's properties to the component's function} (e.g., rubber, because it seals against leaks).

In this context, we test whether \textbf{retrieval-augmented generation (RAG)} \citep{lewis-etal-2020-rag,gao-etal-2024-survey} leads to improvements over standard LLM-based methods. Here, our goal is to bypass standard fine-tuning methods for costly domain-specific adaptation. 
RAG takes user input, applies standard information retrieval techniques to find relevant textual information from an extensive 
database, and leverages retrieved snippets to create a prompt for an LLM. This way, domain-specific information is encoded in the prompt and does not need to be learned through fine-tuning. 
However, LLM-based retrieval may be affected by hallucinations, i.e., semantically valid text providing a plausible interpretation which
contains misleading or \textit{fact-conflicting} information \cite{zhang-etal-2025-sirens}. 
In more detail, we implement and evaluate three approaches include a no-RAG method using LLMs, a method using standard RAG \cite{karpukhin-etal-2020-dense}, and an iterative RAG approach using Chain-of-Verification (CoVe) \cite{ji-etal-2023-towards,press-etal-2023-measuring,dhuliawala-etal-2024-chain}. Following prior work that draws on Wikipedia as a knowledge source for this task \citep[e.g.,][]{eichel-etal-2023-made}, we use Wikipedia as our retrieval corpus.
To assess system performance, we develop a custom full-stack annotation tool that supports crucial functions for evaluating open-ended natural language generation through domain experts.

We define \textbf{plausibility} so as to maximize prediction and explanation quality in a realistic setting with limited resources. Firstly, a predicted \textit{material} is considered plausible for a component if a realistic, domain-specific variant of that component could be made, wholly or partly, of that material \citep{kauf2023event,eichel-etal-2023-made}. This is a permissive, per-material criterion: the material does not need to be the most frequent one or the only option \citep{wang-etal-2018-modeling,porada-etal-2021-modeling}. Secondly, an \textit{explanation} is considered plausible if its rationale for using the material is consistent with established world and domain knowledge, i.e., it contains no fact-conflicting statements \citep{zhang-etal-2025-sirens}. Materials and explanations are judged independently. Following prior work \citep{wang-etal-2018-modeling,eichel-schulte-im-walde-2023-dataset,kauf2023event}, we thus use semantic plausibility as a lens that covers typical materials as well as less typical yet plausible ones,
as grounded in expert knowledge and not 
fluency or coherence.

Our findings are threefold. First, an off-the-shelf LLM predicts plausible materials well, and outperforms prior work. Second, RAG and CoVe do not improve on a No-RAG LLM in our setup: all variants reach similar plausibility scores, while annotators rate No-RAG generations slightly higher, and on average they rank them better. Third, expert agreement is reasonable for plausibility judgments but low for error highlighting and star ratings. 

Our contributions are summarized as follows:
\begin{itemize}[itemsep=0.5pt, topsep=2pt, leftmargin=*]
\item We show that for the task of predicting plausible material candidates for vehicle components,
an \textbf{off-the-shelf LLM-based approach outperforms previous work, without extensive training or fine-tuning}.
\item We investigate explanations for
\textit{why} a specific material candidate is predicted, and reveal that \textbf{RAG-based approaches do not lead to substantial improvements over standard LLM-based methods} in our experiments.
\item We develop a \textbf{custom full-stack annotation tool for expert evaluation} that supports web-based access without installation, secure token-based authentication, automatic progress saving, text-span highlighting, star ratings, 
and drag-and-drop ranking focusing on survey-specific re-usability.\footnote{The annotation tool including documentation of adaptation for a new study is available here: \url{https://anonymous.4open.science/r/annotation-tool-07E3}}
\end{itemize}

\section{Related Work} \label{sec:related-work}

\paragraph{LLMs and RAG for Materials Science} 
While NLP methods have been leveraged for highly specialized domains, LLMs offer promising possibilities for automating material science tasks such as information retrieval, knowledge organization, and innovation derivation and generation \citep{olivetti2020}. Recent work uses LLMs at various stages, including pre-training \citep{kim-etal-2024-melt,oh-etal-2025-incorporating}, dataset creation and verification \citep{song-etal-2023-honeybee}, and modeling \citep{cheung-etal-2024-polyie,jansen-etal-2025-matter}. However, predicting plausible material candidates for the (vehicle) repair domain remains underexplored, in particular when also explaining why a material is predicted. Our work addresses this gap by analyzing LLMs' zero- and few-shot settings for the task at hand.

In addition, we investigate RAG, which combines generation with a retrieval
step.  
\citet{gao-etal-2024-survey} provide a comprehensive survey,
identifying two primary retrieval approaches: \textit{sparse retrieval}
(e.g., BM25 \citep{robertson-walker-1994} based on word overlap) and \textit{dense retrieval} (e.g., DPR \citep{karpukhin-etal-2020-dense}
using neural embeddings). \textit{Hybrid retrieval} combines both retrieval methods \citep{gao-etal-2021-hybrid}. 
RAG approaches for material science tasks such as knowledge extraction and analysis include, among other work, \citep{nidishree-etal-2025} who use embeddings fine-tuned for material science (MatSci-BERT) \citep{gupta2022} and a dense retriever. Other researchers focus on specific materials in a specialized domain, e.g. nano-structured materials
leveraging multilingual embeddings alongside a dense retrieval module \citep{krotkov2025nanostructured}. Our work also focuses on a particular domain and task, however, we employ hybrid retrieval, using \citet{karpukhin-etal-2020-dense}'s embeddings for DPR which we find to substantially outperform MatSci-BERT embeddings. Wikipedia is a standard retrieval corpus in this line of work \citep{lewis-etal-2020-rag, karpukhin-etal-2020-dense}. We further build on prior work \citep{eichel-etal-2023-made} that indicated Wikipedia's usefulness for material prediction in a specialized domain.

\paragraph{Semantic Plausibility and Hallucination} 
Semantic plausibility asks
whether a described state of affairs is plausible given world knowledge, 
which requires more than surface co-occurrence \citep{wang-etal-2018-modeling}. Like prior work on vehicle components \citep{eichel-etal-2023-made}, we target both highly plausible and less frequent but still plausible materials. Plausibility, however, does not aim at providing a truth value: LLMs readily produce fluent output that sounds plausible but conflicts with world knowledge (fact-conflicting) or the given context (unfaithful), a phenomenon known as \textit{hallucination} \citep{zhang-etal-2025-sirens}.\footnote{We refer to \citet{zhang-etal-2025-sirens} for a detailed overview.} A plausible material can thus come with a fact-conflicting explanation, so we assess material and explanation plausibility separately. Prior work mitigates hallucinations in different ways: \citet{ji-etal-2023-towards} reduce hallucinations via self-reflection and \citet{press-etal-2023-measuring} via question decomposition. \citet{dhuliawala-etal-2024-chain} combine both in Chain-of-Verification (CoVe), where the model answers self-generated verification questions and refines its answer. CoVe has been coupled with RAG for computer-assisted design \citep{joseph2025reducing}; we apply it to material prediction with expert evaluation. As our prompts do not ask for citations of retrieved passages, we assess fact-conflicting statements, not faithfulness.


\section{Data} \label{sec:data}
\subsection{Vehicle Components}
\paragraph{Vehicle Component Dataset} 
As targets for our components, we rely on a set of 7,069 unique component names curated by experts from the vehicle repair domain.\footnote{The dataset is provided by a company disclosed upon acceptance.} A component name may denote a tangible physical component such as \textit{cooling blower}, or intangible functional and software components such as \textit{ABS warning lamp function}. 
The dataset comprises 155 single-word 
and 6,914 multiword components with up to eight constituents. 

\paragraph{Evaluation Dataset} To evaluate predicted materials in an expert study, we follow prior work \citep{schlipf:22,eichel-etal-2023-made} and create an evaluation set comprising 100 components 
which focuses on physical (\textit{brake disc}, \textit{spark plug}) vs. software components (\textit{parking assistant}, \textit{catalytic converter monitoring}). We discard physical components that act as a system 
, e.g., \textit{clutch} consisting of the sub-components \textit{clutch pedal}, \textit{clutch disk}, and \textit{master cylinder}. To achieve maximum component variety, we discard components that are highly similar to each other but used in different places in a vehicle, e.g. \textit{pressure control module high-pressure solenoid valve} and \textit{inlet camshaft control solenoid valve}. Here, a high-pressure solenoid valve might need to be constructed from sturdier materials than a solenoid valve dealing with lower pressure but possibly more time-sensitive tasks. 
Whereas prior work relies on 30\% simple single-constituent components, we focus on the distribution of the full dataset mainly consisting of MWEs (98\%), and leave the remaining 2\% of single-word constituents out of consideration.
The resulting set is carefully sampled from real-world data, making its evaluation a valuable contribution to the vehicle repair domain, where comparable research remains scarce.

\subsection{Retrieval Data Sources}
We build the retrieval corpus from Wikipedia because it has been an effective retrieval source in prior work \citep{petroni-etal-2021-kilt,semnani-etal-2023-wikichat,eichel-etal-2023-made}. 
A more informative domain corpus that is also legally usable for research was not available to us.
We develop a custom extraction tool in Rust 
to pull full articles meeting two constraints: (1) articles containing at least two multi-word vehicle components, and (2) articles co-mentioning ''material/materials'' and automotive-domain terms at least three times each. This results in 29,425 and 2,943 articles, respectively, yielding a corpus of 32,368 articles (avg. article length: 4,233 words).

\section{Experiment I: Predicting Plausible Materials for Vehicle Components} \label{sec:exp1}

\paragraph{Modeling} We first perform a baseline experiment to explore LLM performance for plausible material prediction.
We recall that a material is considered \textit{plausible} for a component if a realistic domain-specific variant of that component could be made, wholly or partly, of that material. 
For this, we evaluate the open-source LLM \texttt{Mixtral-8x22B-Instruct-v0.1} \citep{mistral2024mixtral8x22b}.\footnote{We are aware that a wide range of models, including more recent ones, exist and that alternative models may yield different results. Since our work explores relative differences between models, we nevertheless believe that our findings provide valuable insights into a highly underexplored topic.}
We prompt the model in both zero-shot and few-shot settings (cf. App.~\ref{sec:app-exp1}). 
An initial analysis of the output finds relevant candidates in both settings\footnote{One author performed a manual evaluation.}, with few-shot results being more specific than zero-shot results (e.g., \textit{ABS} or \textit{PVC} instead of \textit{plastic}). 
Thus, expert evaluation focuses on few-shot prompting output only.

\paragraph{Evaluation Study} We recruit three annotators who are engineering experts via Prolific to evaluate whether each predicted material is plausible for a given component. We use Google Forms to present five material candidates in a multiple-choice question setup including the option that no material is plausible. Further, annotators could indicate that they do not know the answer to make sure that collected responses are trustworthy. Following prior work \cite{schlipf:22,eichel-etal-2023-made}, we calculate inter-annotator agreement as follows. Given two sets of annotations $A$ and $B$ for the same component, each set encodes the annotator’s choice of plausibility for each material. $a_{i}$ or $b_{i}$ denote ratings of the $i$-th element in the set, and with $\delta_{i}$: 
\begin{equation} \label{eq:iaa}
\delta_i =
\begin{cases}
1 & a_i = b_i \\
0 & a_i \ne b_i
\end{cases}
\end{equation} 
Subsequently, inter-annotator agreement (IAA) is calculated as laid out in Eq.~(\ref{eq:iaa2}): 
\begin{equation} \label{eq:iaa2}
\text{IAA} = \frac{\sum_{i=1}^{|A|} \delta_i}{|A|}
\end{equation} 
IAA values of 1 and 0 denote perfect agreement and disagreement, respectively. Annotator-specific results are shown in App.~\ref{sec:app-exp1}, Table~\ref{tab:annotation-stats}, yielding an average IAA$=0.66$, 95\% [0.64, 0.70], consistent with prior work.

\paragraph{Evaluation Metrics} We use the following metrics
to compare model performance.
\begin{itemize}[itemsep=0.5pt, topsep=2pt, leftmargin=*]
    \item \textsc{Coverage}@$n$: Proportion of components for which at least one material among a model's top-5 predictions is rated plausible by $n$ annotators.
    \item \textsc{Precision}@$n$: Proportion of material predictions among a model's top-5 predictions rated plausible by $n$ annotators.
\end{itemize}

\begin{table}[t]
\centering
\footnotesize
\setlength{\tabcolsep}{3pt}
\begin{tabular}{@{}lccc@{\hspace{8pt}}ccc@{}}
\toprule
 & \multicolumn{3}{c}{Coverage (\%)} & \multicolumn{3}{c}{Precision (\%)} \\
\cmidrule(lr){2-4}\cmidrule(lr){5-7}
 & @1 & @2 & @3 & @1 & @2 & @3 \\
\midrule
\citet{schlipf:22}            & 73.0 & --  & 40.0 & 45.0 & --   & 14.0 \\
\citet{eichel-etal-2023-made} & 93.0 & --  & 73.0 & 62.0 & --   & 28.0 \\
\midrule
Ours & \textbf{100.0} & \textbf{100.0} & \textbf{97.0} & \textbf{96.0} & \textbf{82.4} & \textbf{46.0} \\
\bottomrule
\end{tabular}
\caption{Few-shot LLM results for plausible material prediction, based on the three expert annotators. For results with CI intervals, cf. App~\ref{sec:app-results}. Table~\ref{tab:baseline-with-ci}; 
@2 values denoting majority votes: 100\% coverage, 82.4\% precision.}
\label{tab:baseline}
\vspace{-0.5cm}
\end{table}

\paragraph{Results} 
Table~1 shows that an off-the-shelf LLM performs well: (i) for 96\% [93.0, 100.0] of 100 components, it predicted at least one material rated plausible by all annotators (\textsc{Coverage}), and (ii) 82\% [79.2, 85.4] of predictions rated plausible by at least two of three annotators (\textsc{Precision}). Compared to prior work (@1 and @3), auto-regressive LLMs show a clear gap over domain-adapted encoder-only models such as RoBERTa (DOMAIN RB) \citep{eichel-etal-2023-made} and pattern-based Basilisk \citep{schlipf:22}.

Three differences between the setups qualify this comparison. (i) We built a new evaluation set to remedy key limitations of prior work (cf. \S\ref{sec:data}), so the exact component sets differ. (ii) Our setup permits MWEs as model outputs rather than single constituents, which can increase specificity and reduce ambiguity. (iii) 
The models differ in capabilities and task settings, which we treat as the object of comparison rather than a confound, assessing how a general-purpose LLM fares against prior approaches. 
Beyond these differences, we closely follow the evaluation protocol of \citet{eichel-etal-2023-made}.

In sum, plausible materials can be elicited from LLMs with fair confidence, but it remains unclear \textit{why} a material is predicted. We therefore next ask for an \textit{explanation}, i.e., \textit{a brief rationale linking the material's properties to the component's function}.

\section{Experiment II: Explaining Plausible Material Candidates} \label{sec:exp2}

\subsection{Modeling Approaches}\label{subsec:modeling-approaches}
In this experiment, we explore No-RAG, RAG, and CoVe methods to generate both plausible materials and explanations. We recall that an explanation is considered plausible if its rationale is consistent with established world and domain knowledge, i.e., it contains no fact-conflicting statements. 

\paragraph{No-RAG (Standard Generative QA)} 
Since LLMs are trained on vast, multi-domain data, they store knowledge which prompting can extract without fine-tuning or RAG (Experiment I, \S\ref{sec:exp1}).\footnote{No-RAG partially overlaps with the Experiment I baseline but additionally requests an explanation per material, allowing us to check performance under the extended prompt.} We use a zero-shot question inspired by \citet{zhang-etal-2024-retrievalqa}: 
``\textit{In the automotive context, which materials is the component '<component>' made out of?}'' (full prompt in App.~\ref{sec:app-exp2}, Figure~\ref{box:no-rag-prompt}). The model lists materials sorted by prevalence with a brief explanation each, and suggests plausible materials if it does not know the answer directly. Both LLMs receive the same prompt. This approach relies entirely on knowledge encoded in the model's weights and serves as the reference for RAG and CoVe.

\paragraph{Retrieval-Augmented Generation (RAG)}
We use RAG to augment the standard generation QA prompt with up to ten domain-specific passages before generating the response, 
implementing a hybrid retrieval pipeline.

\begin{figure*}[!t]
    \centering
    \includegraphics[width=0.7\linewidth]{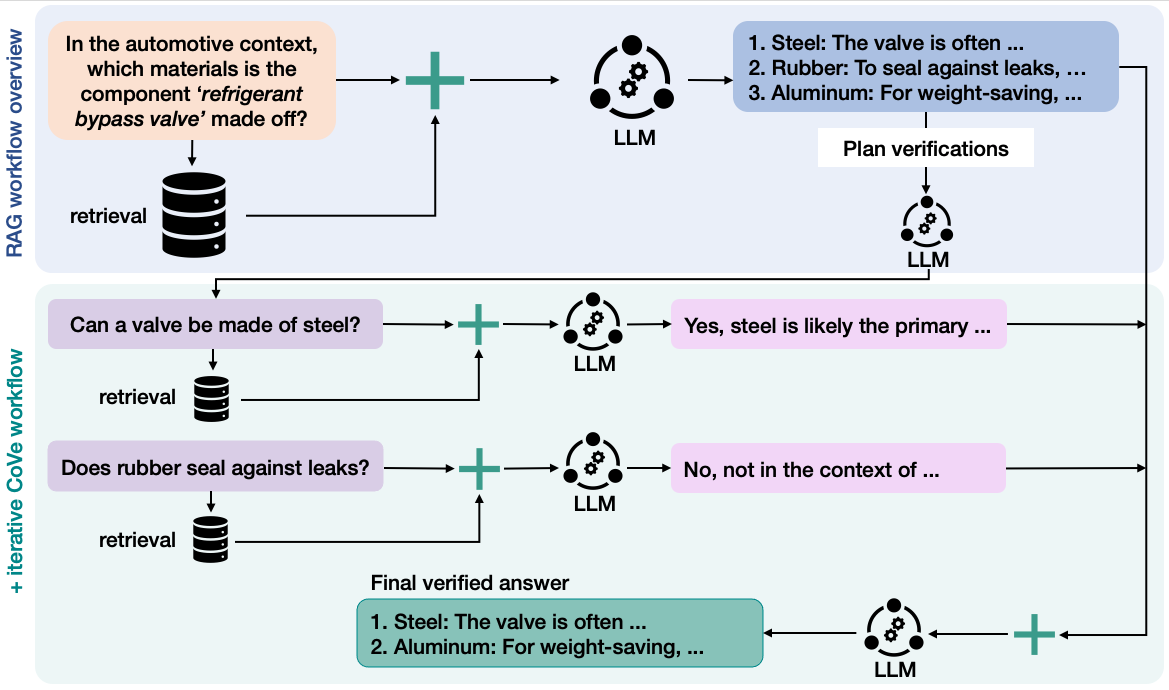}
    \caption{Simplified overviews of (i) standard RAG workflow including the retriever module and a generation model (top, blue box), and (ii) CoVe workflow including a baseline response, planning and execution of verifications, and the final verified answer (bottom, green box).}
    \label{fig:rag-cove}
    \vspace{-0.5cm}
\end{figure*}

\textbf{Retrieval Module}
The retrieval system uses the Haystack framework with OpenSearch as \textit{document store} (chosen over an in-memory store for performance reasons). Ingestion comprises four stages: (1) \textbf{converting} MediaWiki markup to plain text with a patched version of \texttt{wikiextractor}\footnote{\url{https://github.com/attardi/wikiextractor}};  (2)~\textbf{splitting} articles into 200-word passages with a 50-word sliding-window overlap; (3) \textbf{embedding} each passage with the DPR context encoder \citep{karpukhin-etal-2020-dense}
; and (4) indexing passage text and embeddings into OpenSearch, yielding 458,972 passages.

The retriever combines sparse retrieval (BM25, with the component name as query) and dense retrieval (DPR embeddings, with a full question as query). The two paths receive different queries, since sparse retrieval works best with keywords, whereas the dense encoder is trained on whole sentences. Each retriever returns the top-50 passages, which a document joiner merges and re-ranks into the final top-10. 
A manual evaluation of the first 10 components found that roughly 22\% of the retrieved passages 
contained helpful information.

Conversely, about 78\% of the retrieved passages 
were not helpful, which limits what RAG can add (see \S\ref{subsec:discussion}).  

\textbf{Full RAG Workflow} 
To generate a prediction and explanation, the retriever is coupled with an LLM (Figure~\ref{fig:rag-cove}). The dense query is a component-specific question, e.g., ``\textit{What materials does the vehicle component '<component>' consist of?}''. The top-10 passages are prepended to the prompt with instructions to answer even if materials are not mentioned directly and to avoid citing the full passages in the output.

\paragraph{Chain-of-Verification (CoVe)}
CoVe \citep{dhuliawala-etal-2024-chain} 
extends RAG with an iterative self-verification loop to reduce hallucinations in the generated answers. The process includes the following steps visualized in Figure~\ref{fig:rag-cove}.

\begin{enumerate}[itemsep=0pt, topsep=2pt, leftmargin=*]
\item \textbf{Generate Baseline Response}: The model generates an initial material list with explanations (identical to the RAG approach above).
\item \textbf{Plan Verifications}: The model receives the original question and initial response, then generates a set of targeted verification questions. 
\item \textbf{Execute Verifications}: Each verification question is answered independently using the retrieval pipeline, which can surface information about materials not retrieved in Step 1.
\item \textbf{Generate Final Verified Response}: The initial response and all verification QA pairs are concatenated into a final prompt, and the model refines its answer accordingly.
\end{enumerate}

The key advantage of CoVe is that verification questions can uncover new retrieval targets. For example, the initial answer for \textit{engine piston} may list typical piston materials together with the candidate \textit{tin}. Unless the model learned about tin in piston manufacturing during training, it cannot tell that tin's melting point is too low. With CoVe, the model can generate a verification question such as ``\textit{Is tin a suitable material for engine pistons?}''. This prompts the retriever to fetch information about tin and allows the model to conclude that tin is not a plausible piston material. For a full prompt example, see App.~\ref{sec:app-exp2}, Figure~\ref{box:cove-prompt}.

\paragraph{Experimental Setup} All three methods use both \texttt{Mixtral-8x22B-Instruct-v0.1}  \cite{mistral2024mixtral8x22b} and \texttt{GPT-4o} \cite{openai2024gpt4ocard} (cf. App.~\ref{sec:app-exp2} for details), yielding six output variants per component.

\subsection{Expert Evaluation Study} \label{sec:evaluation-study}

\paragraph{Study Design}
We 
conduct a human expert evaluation study to evaluate No-RAG and RAG-based system performance. Evaluation was divided along three aspects, following prior work \cite{zhong-etal-2022-towards}, which are elicited through five tasks. 

\paragraph{Correctness} As LLMs are prone to hallucination, annotators are instructed to determine whether a material prediction is plausible for a component. They need to further verify whether an explanation is not consistent with their expert knowledge.
\begin{itemize}[itemsep=0pt, topsep=2pt, leftmargin=*]
    \item \textit{Task 1}: Annotators are shown a list with all material predictions, and are asked to select all plausible materials.
    \item \textit{Task 2}: Annotators are shown the full output, including explanations. They are asked to highlight all fact-conflicting parts.
\end{itemize}
\paragraph{Completeness} Annotators assess whether material predictions and explanations are complete, ensuring that all expected materials are included, and explanations provide sufficient information.\begin{itemize}[itemsep=0pt, topsep=2pt, leftmargin=*]
\item \textit{Task 3}: Annotators are tasked with rating the quality of the material predictions on a scale of one to five where 5 denotes the best option. If they think a material is missing, they are asked to enter it into an optional comment text field.
\item \textit{Task 4}: Similar to the previous task, annotators are tasked with rating the quality of the explanations on a scale of one to five. They can enter which information they miss or which is superfluous in a text field.
\end{itemize}
\paragraph{Ranking} Annotators are asked to rank the outputs according to their subjective expert preferences. Importantly, they should only consider the contents of the output when ranking, not the formatting or style.
\begin{itemize}[itemsep=0pt, topsep=2pt, leftmargin=*]
\item \textit{Task 5}: For each component, annotators are shown all six outputs simultaneously. They are asked to rank the outputs based on their preference, as shown in Figure~\ref{fig:wagner-ui-2}.
\end{itemize}
Tasks 1--4 are displayed simultaneously; Task 5 is shown individually (cf. App~\ref{sec:app-evaluation-study}, Figures~\ref{fig:wagner-ui-1} and \ref{fig:wagner-ui-2}).

\paragraph{Annotation Tool}
We develop a custom full-stack annotation tool from scratch since no existing open-source or cost-free annotation tool satisfied all requirements for a suitable evaluation platform.\footnote{Since Google Forms was used in the baseline study, it would be an obvious choice. However, the tool lacks highlighting support and does not offer an annotator-friendly way to implement the ranking task. Potato \citep{pei-etal-2022-potato} has limited security and documentation, and LimeSurvey's JavaScript customization pathway is is relatively unexplored with no working community examples.} The tool is built using TypeScript, NestJS (backend), SQLite (database), and Vue with PrimeVue (frontend). It supports web-based access without installation, secure token-based authentication, automatic progress saving, text-span highlighting for marking factual errors, star ratings, optional free-text comments for missing materials, and user-friendly drag-and-drop ranking of multiple outputs simultaneously. Our tool is designed as a reusable boilerplate, with survey-specific logic implemented as extensions to the core framework.

\begin{table}[t]
\centering
\small
\begin{tabular}{lrrrr}
\toprule
 & \multicolumn{2}{c}{Coverage@2 (\%)} & \multicolumn{2}{c}{Precision@2 (\%)} \\
\cmidrule(lr){2-3}\cmidrule(lr){4-5}
 & Mixtral & GPT-4o & Mixtral & GPT-4o \\
\midrule
No-RAG & 100.0 & 98.0 & 87.3 & 89.7 \\
RAG    & 100.0 & 92.0 & 88.0 & 93.7  \\
CoVe   & 100.0 & 98.0 & 89.3 & 91.2  \\
\bottomrule
\end{tabular}
\caption{Overview of results for plausible material prediction (Task 1), comparing 
No-RAG, RAG, and CoVe approaches. Metrics are calculated at majority vote (@2 out of 3 expert annotators).}
\label{tab:task1}
\vspace{-0.5cm}
\end{table}

\paragraph{Setup and Participants}
We evaluate 50 components (5 of 10 batches of 10 components each), and recruit three annotators per batch via Prolific. Participants are pre-screened to be located in the US, UK, or Germany and fluent in English; two hold engineering degrees and one a materials science degree per batch.\footnote{Imbalance in expert background stems from a surplus of annotators with an engineering vs. materials science degree.}
Participants are guided through an initial interactive tutorial explaining how the tool works and receive detailed introduction to all the tasks. We embed attention checks in each batch; one participant who failed both checks was excluded per Prolific guidelines. For further details on the study setup, we refer to App.~\ref{sec:app-evaluation-study}.

\paragraph{Annotator (Dis)agreement}
To determine annotation reliability, we calculate agreement among annotators leveraging the following metrics for the various tasks.
For Task 1 (plausibility selection), agreement is reasonable: IAA (Eq.~\ref{eq:iaa2}) is 0.77 (95\% bootstrap CI over components [0.75, 0.79]) vs. 0.66 in the baseline study (cf. Table~\ref{tab:baseline}). We report Task 1 at majority vote.
For Task~2 (highlighting), not enough overlap between annotators is observed
to allow for the calculation of a meaningful metric. We therefore interpret the error rates in Table~\ref{tab:domain-rag-factual-errors-ranking} as errors noticed by individual annotators rather than verified rates (cf. App.~\ref{sec:app-exp2-agreement} for details).
For Tasks 3 and 4 (star ratings), absolute agreement is below chance (Krippendorff's $\alpha=-0.15$ for materials, $-0.17$ for explanations; App.~\ref{sec:app-exp2-agreement} for details), reflecting differences in annotator strictness. Since the same annotators rated all variants of a component, such offsets cancel in paired differences, which we quantify with bootstrap CIs.
For Task~5 (ranking), agreement is weak to moderate (mean pairwise Kendall's $\tau=0.24$ 
\citep{kendall-1938} and may be inflated by the shared initial order, which was randomized per component and does not favor any method.
Overall, agreement varies across tasks. However, as annotators judged all variants of each component, methods can still be compared on all tasks despite lower absolute agreement for Tasks 2--4. We discuss potential sources for this in §\ref{sec:results} and §\ref{sec:limitations}.

\begin{table}[t]
\centering
\small
\begin{tabular}{l|rr}
\toprule
\multicolumn{1}{c}{}         & Mixtral & GPT-4o   \\
\midrule
No RAG   & 22\%    & 22\%       \\  
RAG      & 30\%    & 18\%    \\
CoVe     & 28\%    & 18\%        \\
\bottomrule
\end{tabular}
\caption{Overview of model outputs flagged to contain at least one fact-conflicting statement.}
\label{tab:domain-rag-factual-errors-ranking}
\vspace{-0.5cm}
\end{table}

\begin{table*}[t]
\centering
\small
\begin{tabular}{lccc cc cc}
\toprule
 & \multicolumn{3}{c}{\textbf{Mean (both models)}} & \multicolumn{2}{c}{\textbf{RAG $-$ No-RAG}} & \multicolumn{2}{c}{\textbf{CoVe $-$ No-RAG}} \\
\cmidrule(lr){2-4}\cmidrule(lr){5-6}\cmidrule(lr){7-8}
 & No-RAG & RAG & CoVe & $\Delta$ [95\% CI] & $p_{\text{adj}}$ & $\Delta$ [95\% CI] & $p_{\text{adj}}$ \\
\midrule
Material rating ($\uparrow$) & 4.26 & 4.02 & 4.12 & $-0.24$ [$-0.38$, $-0.12$] & $<0.001$ & $-0.14$ [$-0.25$, $-0.04$] & 0.03 \\
Explanation rat. ($\uparrow$) & 4.18 & 3.88 & 4.08 & $-0.30$ [$-0.44$, $-0.18$] & $<0.001$ & $-0.11$ [$-0.21$, $-0.01$] & 0.06 \\
Rank position ($\downarrow$) & 2.18 & 2.71 & 2.61 & $+0.53$ [$+0.14$, $+0.93$] & 0.03 & $+0.43$ [$+0.04$, $+0.82$] & 0.06 \\
\bottomrule
\end{tabular}
\caption{Results and differences between methods on rating (rat.) and ranking tasks (Tasks 3--5). $\Delta$ is the paired mean difference over the 50 components (annotator means, both models averaged), with a 95\% bootstrap CI (50,000 resamples of components). $p_{\text{adj}}$ are bootstrap $p$-values, Holm-corrected across the six comparisons in this table. Rank 0 is best and 5 is worst, so lower is better for ranks.} 
\label{tab:primary}
\vspace{-0.3cm}
\end{table*}

\subsection{Results} \label{sec:results}

\paragraph{Task 1: Plausible Material Prediction}
Results for the prediction of plausible material candidates are shown in Table~\ref{tab:task1}, using \textsc{coverage} and \textsc{precision} at majority vote.\footnote{We use majority vote to account for varying numbers of annotators. We refer the reader for results for \textsc{precision}@3 and \textsc{coverage}@3 and a detailed discussion to App.~\ref{sec:app-results}.} While overall performance is strong, we next examine whether the observed differences are statistically significant. 
As shown in App~\ref{sec:app-results}, Table~\ref{tab:task1-sign-testing}, RAG and CoVe did not change how often materials were judged plausible (OR 1.16, 95\% CI [0.94, 1.43] and 1.08 [0.88, 1.33]) when compared to No-RAG. There is no sign that this differs between the two models (interaction $p=.67$). At model level, GPT-4o predictions were judged slightly more often than Mixtral's but the difference is small (OR 1.28, [1.05, 1.56], $p=0.014$, unadjusted).

We further analyze model performance regarding materials by prevalence. Here, we expect most used and thus highly plausible candidates to be listed first. Less used and thus less plausible materials should be listed last. Results are shown in App.~\ref{sec:app-results}, Table~\ref{tab:domain-rag-first-last}, indicating that the models indeed sorted materials as instructed: first predictions were on average more often rated plausible (88.9\%) than last predictions (85.1\%).

\paragraph{Task 2: Fact-Conflicting Statement Detection}
Task~2 asks annotators to highlight fact-conflicting sections (hallucinations) in predictions and explanations. Of 177 highlighted sections, 85 remain after manually removing accidental highlights (28), redundant markings of materials already rejected in Task~1 (51), statements that are actually correct (4), and invalid outputs (9). Only two of them were flagged by more than one annotator, so the results reflect errors noticed by individual annotators, not agreed-upon errors. Table~\ref{tab:domain-rag-factual-errors-ranking} shows the outputs with at least one flagged error. Methods show no consistent difference (RAG and CoVe are flagged more often than No-RAG for Mixtral, but not for GPT-4o), while GPT-4o is flagged less often than Mixtral (29 vs.\ 40 of 150 outputs). Of the 85 sections, 40 concern materials rated plausible in Task~1 and 45 implausible ones (App.~\ref{sec:app-results}), so a flag may target an implausible material and not an error in the explanation, and the rates may overestimate actual errors. Flagged errors show no tendency to occur more often in the first or last predicted material.

\paragraph{Task 3--5: Quality and Preference Ratings}
The goal of these three tasks is to understand whether material predictions (Task 3) and explanations (Task 4) do not lack relevant materials or contain superfluous information. Further, subjective preferences of expert annotators are elicited (Task 5).\footnote{A potential confound is the default order randomized per component but not per annotator: if annotators were biased toward minimal reordering, this could inflate agreement and reduce sensitivity to actual differences.} 
To assess whether one method or model performs better than the others, we compare them with paired confidence intervals, reported in Table~\ref{tab:primary}. 
Compared to No-RAG, RAG received lower material ratings ($\Delta=-0.24$, 95\% CI $[-0.38,-0.12]$) and explanation ratings ($\Delta=-0.30$, $[-0.43,-0.18]$) and was ranked worse ($\Delta=+0.53$ positions, $[+0.13,+0.93]$). CoVe showed smaller deficits whose intervals only just excluded zero, and was rated higher than RAG for explanations ($\Delta=+0.20$, $[+0.10,+0.30]$). This is in contrast to Task-1 results where RAG and CoVe performance does not differ. Regarding model performance, findings across tasks agree: an exploratory analysis of the interaction between method and model shown in App.~\ref{sec:app-results}, Table~\ref{tab:exploratory} did not suggest a detectable difference between GPT-4o and Mixtral.  

\subsection{Ablation experiment} \label{subsec:discussion}
While prior work found Wikipedia to be an effective retrieval source, RAG did not improve over No-RAG in our study. To investigate this finding, we analyze retrieved passages for direct material evidence and noise. 
We classify each retrieved passage as (i) \textit (naming the component and a material within a 2-sentence window), (ii) \textit, (iii) \textit (distractor: materials of something else), or (iv) \textit (neither). This is stricter than the manual analysis in \ref{subsec:modeling-approaches} which also counted passages supporting indirect reasoning (e.g., refrigerants implying rubber seals). For full details, cf. App~\ref{sec:app-ablation}.
Among the top-10 passages retrieved from the 32,368-article Wikipedia subset, 11.7\% (95\% CI [9.1, 14.4], Mixtral) and 13.0\% (95\% CI [10.1, 16.0], GPT-4o) contained direct evidence
(cf.  Table~\ref{tab:retrieval-analysis} for full results). As this analysis is descriptive, note that we cannot conclude that (i) Wikipedia lacks relevant evidence or (ii) the retrieved passages caused lower ratings. These figures rather represent retrieval lower-bound estimates. 
Nevertheless, they highlight the limitations of Wikipedia as a retrieval source in our setup.

\section{Conclusion} \label{sec:conclusion}
We addressed the prediction of plausible materials for vehicle components. A baseline experiment showed that a simple LLM approach compares favorably to prior work. In our main study, we extended the task to require explanations and compared No-RAG, standard RAG, and Chain-of-Verification (CoVe) across two established LLMs. An expert evaluation with a custom-built annotation tool yielded mixed results. RAG and CoVe did not improve plausibility over No-RAG and were partly rated and ranked lower. Wikipedia retrieval helped in prior work but not in our setup. 
Despite these limitations, the results are encouraging as LLM-based systems hold promise as expert decision support for vehicle repair and related industrial applications.

\section*{Limitations} \label{sec:limitations}

\paragraph{No gold standard, small evaluation sets.}
In the automotive domain, there is a general lack of comparable research, so our study already at this stage provides a valuable novel contribution. While we acknowledge the limited size of the evaluation dataset, we ground our work in carefully sampled data points and expert human judgments. Moreover, we explicitly address the scale of the experiments so as to avoid unwarranted implications of generalizability. 
Specifically, the lack of an annotated gold standard means that automatic metrics cannot be used and every system variant must be judged by human experts. This limits scale: we evaluate 100 components (Experiment I) and 50 components (Experiment II), and a full evaluation per hyperparameter setting (top-$k$, passage length, prompts) was infeasible. These settings were chosen through targeted manual checks, not tuned systematically. 
We thus interpret differences between methods with paired bootstrap confidence intervals over components.

\paragraph{Expert annotation.}
Agreement is reasonable for plausibility judgments (Exp. I: 0.66, Exp II: 0.77), lower for error highlighting (Fleiss' $\kappa=0.11$, AC1 $=0.59$), and below chance for star ratings ($\alpha<0$). Plausible reasons include the deliberately open-ended questions, which leave room for domain-specific interpretation (e.g., ``\textit{plastic is commonly used in car interiors}'' is judged differently for luxury and budget cars), and different backgrounds of crowdsourced experts. Some annotators finished well below the estimated 65 minutes, and the platform's payment model may reward speed over care. With three annotators per batch, our intervals capture variation across components but not across annotators. We report all tasks regardless of agreement and base method comparisons on paired differences, since the same annotators judged all variants of a component.

\paragraph{Comparability to prior work.} 
The margin over prior work remains large (e.g., \textsc{Precision}@3 of 46\% vs.\ 28\%, Table~\ref{tab:baseline}), although those systems were evaluated using a different evaluation set and different restrictions on model output. In cf. §\ref{sec:exp1}, we specify differences in detail and lay out why the comparison is nevertheless meaningful.

\paragraph{Why RAG did not help remains partly open.}
Potential explanations for the low RAG performance include, among others, noise in the retrieval corpus, a potentially sub-optimal top-$k$, and prompt sensitivity. All of these are hypotheses. Within our setup, it was not feasible to to vary the retriever, $k$, or prompts. 

\paragraph{Data and generalization.}
Component names come from a single English-language dataset, and the evaluation set excludes multi-part systems and near-duplicates (\S\ref{sec:data}). Results may not transfer to other languages, systems, or other repair settings.

\paragraph{Models and prompts.}
We test two models (Mixtral-8x22B-Instruct and GPT-4o) with prompts adapted from related work. Patterns are similar across both, but newer models or tuned prompts may behave differently, and the closed GPT-4o limits reproducibility.

\paragraph{Plausibility is not repair-readiness.}
We evaluate whether materials are plausible and whether annotators notice errors, not whether outputs are correct for a specific vehicle or sufficient to guide a repair. Flagged errors are unverified, and we did not test with mechanics in a deployed setting. The practical promise reported here is therefore suggestive, and outputs should be reviewed by experts (see Ethical Considerations).

\smallskip

These limitations qualify the size and generality of the differences between methods. They do not change the central findings: off-the-shelf LLMs predict plausible materials well (high agreement, large margin over prior work), and in our setup neither RAG nor CoVe improved on them.

\section*{Ethical Considerations}
\paragraph{Intended use, failure modes, and safeguards.}
We position the systems as \emph{expert decision support}, not autonomous repair guidance. Our evaluation targets plausibility and explanation quality. It does not aim at evaluating vehicle-specific correctness or use in practice. Failure modes include undetected errors, confusion with related components, generic answers that overlook variant-specific requirements, irrelevant retrieved passages, and over-trust in fluent explanations. Safety-critical components were not analyzed separately. We recommend, but did not evaluate, safeguards such as verification against authoritative documentation, expert review of safety-critical components, displaying sources and unverified status, and a practitioner study before deployment.

\paragraph{Data and Models}
For the presented RAG approaches, we harness publicly available data. More specifically, we use a portion of the English Wikipedia customized to the domain of interest. We acknowledge that Wikipedia text content including Wikipedia dumps is licensed under both the Creative Commons Attribution-ShareAlike 3.0 License and the GNU Free Documentation License.

We use and adapt an open-source LLM as provided and licensed under the Apache License 2.0 by \texttt{huggingface} \cite{wolf-etal-2020-transformers}. We further use a closed, proprietary model with inaccessible training datasets and algorithmic weights. We acknowledge the possibility of (accidental) breaches of data privacy, systemic model bias, transparency, and reproducibility introduced by the use of such a model. 

\paragraph{Human Expert Evaluation}
In the context of our evaluation tasks, we collected ratings from human participants. Annotation was fully voluntary and could be stopped at any time without providing any reasons. We paid participants fairly according to the platform's recommendation, communicated decisions transparently, and reached out to individual participants whenever necessary during the annotation approval process. We obtained consent for the collected data by using an established platform (Prolific) to carry out the human annotation.

\paragraph{Use of AI Assistants} The authors acknowledge the use of AI assistants for correcting grammatical errors, optimizing coherence and length within selected paragraphs, formatting boxes and tables, and generating generic code snippets for implementing the experiments, the ablation experiment, and significance testing.


\bibliography{custom}

\appendix

\section{Experiment I}
\label{sec:app-exp1}

\begin{tcolorbox}[
    colback=blue!8!white,
    colframe=black,
    arc=4pt,
    boxrule=0.8pt,
    left=8pt, right=8pt, top=6pt, bottom=6pt, fontupper=\small] \label{box:baseline-prompt}
Your task is to predict a comma-separated list of up to five materials that a vehicle
component is made of. For example:\\
\\
Component: brake disc\\
Materials: grey cast iron, carbon-ceramic composite, ceramic\\
\\
Component: motor oil\\
Materials: mineral oil, synthetic oil, additives\\
\\
Component: igniter\\
Materials: nickel, aluminium oxide, sintered alumina, steel\\
\\
Please predict the materials for the following component: {component}\\
Just provide the names of the materials, separated by commas, without any explanation.
Answer:
\end{tcolorbox}

Annotator-specific results are shown in Table~\ref{tab:annotation-stats}. Originally, experiments and results included annotations from one of the authors. The average IAA is 0.69 across all four annotators. As Experiment II uses three expert annotators, we excluded previously included author annotations for a cleaner comparison, resulting in an IAA of 0.66 across the three expert annotators only. In any of the setups, IAA is consistent with prior work \citep{schlipf:22,eichel-etal-2023-made}.

\begin{table*}[t]
\centering
\scriptsize
\begin{tabular}{l|ccc|ccc}
\toprule
\multicolumn{1}{c}{}& \multicolumn{3}{c}{\textsc{Coverage}} & \multicolumn{3}{c}{\textsc{Precision}}  \\
\cmidrule(lr){2-4} \cmidrule(lr){5-7}
             & \multicolumn{1}{c}{@1} & \multicolumn{1}{c}{@2} &\multicolumn{1}{c}{@3} & \multicolumn{1}{|c}{@1} & \multicolumn{1}{c}{@2} &
             \multicolumn{1}{c}{@3} \\
\midrule
\citet{schlipf:22}    & 73\%  & --    & 40\%      & 45\%    &--   & 14\%       \\
\citet{eichel-etal-2023-made}    & 93\%  & --      & 73\%      & 62\%   & --      & 28\%       \\
\midrule
Our approach & \textbf{100\% }[100.0, 100.0] & \textbf{100\% }[100.0, 100.0] &  \textbf{97\%} [93.0, 100.0] & \textbf{96\% }[94.6, 98.0] & \textbf{82.4} [79.2, 85.4] &\textbf{46\%} [41.8, 50.2] \\
\bottomrule
\end{tabular}
\caption{Few-shot LLM results for plausible material prediction, based on the three expert annotators. Brackets show 95\% bootstrap confidence intervals (5,000 resamples of the 100 components); 
@2 denotes a majority vote.}
\label{tab:baseline-with-ci}
\vspace{-0.3cm}
\end{table*}

\begin{table}[t]
\centering
\small
\begin{tabular}{lrrr}
\toprule
 & A1 & A2 & A3 \\
\midrule
A1 & -- & 0.75 & 0.60 \\
A2 & 0.75 & -- & 0.64 \\
A3 & 0.60 & 0.64 & -- \\
\bottomrule
\end{tabular}
\caption{Pairwise inter-annotator agreement (\ref{eq:iaa2}) between the three expert annotators in Experiment~I. 
The average over the three pairs is 0.66.}
\label{tab:annotation-stats}
\end{table} 

\section{Experiment II} \label{sec:app-exp2}

\begin{tcolorbox}[
    colback=blue!8!white,
    colframe=black,
    arc=4pt,
    boxrule=0.8pt,
    left=8pt, right=8pt, top=6pt, bottom=6pt, fontupper=\small] \label{box:no-rag-prompt}
Please answer the question below. When listing multiple materials, sort them by the
amount of the material used, from the most used to the least used.
Please also provide a brief explanation for each material.
If you do not know the answer directly, please suggest plausible materials that
could be used in the component.\\
\\
Question: In the automotive context, which materials is the component ''{component}'' made out of?\\
Answer:
\end{tcolorbox} 

\paragraph{Experimental Setup} Throughout our experiment we use \texttt{Mixtral-8x22B-Instruct-v0.1}  \cite{mistral2024mixtral8x22b} and \texttt{GPT-4o} \cite{openai2024gpt4ocard} as language models. We access \texttt{Mixtral} through \texttt{huggingface} \cite{wolf-etal-2020-transformers}, and run it locally on two NVIDIA RTX A600 GPUS with 4-bit quantization with default parameter settings. We access GPT-4o through the OpenAI API.

For RAG, we use OpenSearch\footnote{\url{https://opensearch.org/}} as document store which supports both fast full-text search via an inverted index (sparse retrieval), and retrieval using vector embeddings of the text (dense retrieval). The OpenSearch instance is hosted on a private server including a two-way authentication with transport layer security (TLS) certificates for secure communication.
An OpenSearch dashboard is deployed to explore the data and test queries directly with a graphical user interface.
To create embeddings, we use a model\footnote{\url{https://huggingface.co/facebook/dpr-ctx_encoder-multiset-base}} by \citet{karpukhin-etal-2020-dense} which we obtain through \texttt{huggingface}.

\subsection{Evaluation Study} \label{sec:app-evaluation-study}

\subsection{Annotation Costs}
Domain expert annotations collected in the context of Experiment I (§\ref{sec:exp1}) and Experiment II (§\ref{sec:exp2})  required a budget of 78\$ and 202\$, respectively. We follow the used platform's guidelines regarding fair payment and compensate completion times exceeding our initial estimates with a bonus payment.

\subsection{Annotator (Dis)Agreement} \label{sec:app-exp2-agreement}

\begin{table}[t]
\centering
\small
\begin{tabular}{lrr}
\toprule
 & Estimate & 95\% CI \\
\midrule
flagged by $\geq 1$A (\%) & 43.7 & [34.0, 53.0] \\
flagged by $\geq 2$A (\%) & 14.0 & [8.7, 20.0] \\
flagged by all 3A (\%) & 1.3 & [0.3, 2.7] \\
\midrule
Pairwise agreement & 0.72 & [0.66, 0.78] \\
Fleiss' $\kappa$ & 0.11 & [0.03, 0.18] \\
Gwet's AC1 & 0.59 & [0.46, 0.70] \\
\bottomrule
\end{tabular}
\caption{Agreement on fact-conflicting flags (Task 2) at output level. An output counts as flagged by an annotator (A) if they highlighted at least one valid span in it. Based on 300 outputs (50 components $\times$ 6 variants) with three annotators each. Intervals are percentile bootstrap intervals over components (5,000 resamples).}
\label{tab:task2-agreement}
\end{table}

\paragraph{Tasks 3 and 4}

We compute Krippendorff's $\alpha$ (ordinal) jointly over the three annotators of each batch, with the 60 outputs of a batch as units, and average over the five batches. Annotators differ across batches, so $\alpha$ is never averaged across them. 95\% confidence intervals are percentile bootstrap intervals. Results are $\alpha=-0.15$ (95\% CI $[-0.22,-0.09]$) for material ratings and $\alpha=-0.17$ ($[-0.24,-0.11]$) for explanation ratings.\\
\textbf{Differences in strictness}: Annotator mean ratings within a batch range from 2.5 to 5.0 stars for materials and from 2.9 to 5.0 for explanations. 
\\
\textbf{Rank agreement}: To separate strictness from ranking, we compute Spearman's $\rho$ between annotator pairs over the outputs of a batch. Of 15 pairs, 13 are defined. 
Values range from $-0.31$ to $0.35$ for materials and from $-0.26$ to $0.27$ for explanations, with means of $0.04$ and $-0.03$. Rank agreement is thus weak but not systematically negative: annotators share little signal about which outputs are better, but do not contradict each other.
\\
\textbf{Implications}: Single ratings carry little information, so we do not interpret differences between individual outputs. Method comparisons rest on paired averages over 50 components, in which annotator strictness cancels because the same annotators rated all variants of a component. Annotators are confounded with batches, so the results describe this group of annotators. 

\subsection{Results} \label{sec:app-results}

\begin{table*}[h]
\centering
\small
\begin{tabular}{lrrrrrr}
\toprule
 & \multicolumn{2}{c}{Coverage@2 (\%)} & \multicolumn{2}{c}{Precision@2 (\%)} & & \\
\cmidrule(lr){2-3}\cmidrule(lr){4-5}
 & Mixtral & GPT-4o & Mixtral & GPT-4o & OR [95\% CI] & $p$ \\
\midrule
No-RAG & 100.0 & 98.0 & 87.3 & 89.7 & ref. & \\
RAG    & 100.0 & 92.0 & 88.0 & 93.7 & 1.16 [0.94, 1.43] & 0.16 \\
CoVe   & 100.0 & 98.0 & 89.3 & 91.2 & 1.08 [0.88, 1.33] & 0.48 \\
\midrule
GPT-4o vs.\ Mixtral   & & & & & 1.28 [1.05, 1.56] & 0.014 \\
Method $\times$ model & & & & & $\chi^2(2)=0.79$ & 0.67 \\
\bottomrule
\end{tabular}
\caption{Plausibility of predicted materials (Task 1). Coverage and precision are majority-vote values (at least two of three annotators). Odds ratios (OR) come from a logistic GEE on individual annotator judgments, clustered by component. No-RAG and Mixtral are used as reference levels. Method effects are averaged over both models. All $p$-values are unadjusted.}
\label{tab:task1-sign-testing}
\end{table*}

\paragraph{Task 1} We assess differences in performance between methods and models using a logistic GEE, with components as clusters and an exchangeable working correlation.
Method, model, and their interaction are used as predictors. Annotator effects are not modeled separately.
We show additional results for Task 1 using metrics in correspondence to Experiment I (§\ref{sec:exp1} in Table~\ref{tab:cov-prec-n}). To further analyze the disagreement within the collected annotations, we calculate IAA per batch. Result are shown in Table~\ref{tab:iaa-per-batch}, indicating that annotator disagreement is particularly low in case of batch 2. 

It should be further noted that given the agreement percentages seen in the different studies, \textsc{coverage} and \textsc{precision} heavily depend on the number of annotators, and are thus not directly comparable. For example, when calculating \textsc{coverage}@3 with 100 annotators, it is highly likely that at least one material suggestion will be deemed plausible by at least three annotators, as this represents just 3\% of the total. In this study, however, three annotators account for 100\% of the evaluations, meaning a material suggestion must be considered plausible by all annotators, which is much less likely. The same applies to \textsc{precision}@3. We thus present results at majority vote in Table~\ref{tab:cov-prec-n}.

\paragraph{Task 2}
We further analyze the relation between the highlighted sections and annotator decisions for Task 1. 40 out of the 85 highlighted sections are part of explanations for materials that are considered plausible by the annotator in the previous task. In contrast, 45 sections are part of explanations relating to material that was considered implausible in the previous task. Here, it remains unclear whether the highlighting of these sections stems from an implausible material candidate or fact-conflicting statements in the explanation. Consequently, the error rate presented in Table~\ref{tab:domain-rag-factual-errors-ranking} may overestimate the number of fact-conflicting errors. 

\section{Ablation Experiment}  \label{sec:app-ablation}

\paragraph{Material lexicon.} The material lexicon decides what counts as a ``material mention'', so the results depend on it. It has two parts:
(i) a non-exhaustive seed list of common automotive materials, and (ii) 
terms extracted from the model answers, kept if they occur in at least >= 2 answers.
Ambiguous words such as lead are excluded. We further exclude a specified set of common non-materials.

Seed material lexicon: \textit{stainless steel; steel; cast iron; grey iron; gray iron; iron; aluminium; aluminum; alumina; copper; brass;
bronze; zinc; nickel; magnesium; titanium; chromium; tungsten; platinum; solder; metal; alloy;
plastic; polymer; thermoplastic; polyamide; nylon; polypropylene; polyethylene; polycarbonate; polystyrene;
polyurethane; polyvinyl chloride; pvc; acrylonitrile butadiene styrene; ptfe; teflon; polyethylene terephthalate;
epoxy; phenolic; resin; rubber; elastomer; silicone; epdm; neoprene; latex; glass; ceramic; silicon carbide;
silicon; quartz; carbon fiber; carbon fibre; graphite; fiberglass; fibreglass; composite; leather; fabric;
textile; wood; cork; foam; high strength steel; advanced high strength steel; dual phase steel; boron steel; carbon steel; mild steel; galvanized steel; galvanised steel; alloy steel; spring steel; tool steel; bearing steel; electrical steel; forged steel; cast steel; ductile iron; nodular iron; malleable iron; compacted graphite iron; wrought iron; cast aluminium; cast aluminum; aluminium alloy; aluminum alloy; magnesium alloy; titanium alloy; zinc alloy; zamak; copper alloy; beryllium copper; phosphor bronze; cupronickel; babbitt; inconel; nickel alloy; tin; lead-free solder; gold; silver; palladium; rhodium; iridium; cobalt; molybdenum; manganese; vanadium; lithium; neodymium; dysprosium; praseodymium; ferrite; alnico;
polyamide 6; polyamide 66; polyamide 12; pa6; pa66; pa12; abs; asa; san; pp; hdpe; ldpe; pet; pbt; polybutylene terephthalate; pom; acetal; polyoxymethylene; pmma; acrylic; pc; pc abs; ppo; polyphenylene oxide; pps; polyphenylene sulfide; peek; ppa; polyphthalamide; pu; tpu; tpo; tpe; tpv; evoh; pvb; polyvinyl butyral; eva; polyester; polyimide; pvdf; fep; pfa; etfe; fluoropolymer; polyolefin; xlpe; glass filled nylon; glass fiber; glass fibre; smc; bmc; gfrp; cfrp; kevlar; aramid; natural fibre composite; flax; kenaf;
nbr; nitrile rubber; hnbr; fkm; viton; acm; aem; butyl rubber; natural rubber; synthetic rubber; sbr; polybutadiene; polyisoprene; chloroprene; csm; bromobutyl; chlorobutyl; fluorosilicone; liquid silicone rubber; silicone rubber; thermoplastic elastomer; polyurethane foam; expanded polypropylene; expanded polystyrene; melamine foam; carbon black; silica; steel cord; tire rubber;
laminated glass; tempered glass; borosilicate glass; glass wool; mineral wool; cordierite; zirconia; ceria; zeolite; titanium dioxide; aluminium oxide; aluminum oxide; silicon nitride; aluminium nitride; porcelain; mica; fused silica; pzt; sapphire; friction material; ceramic fibre; expanded graphite; synthetic graphite; natural graphite;
lithium iron phosphate; nickel manganese cobalt; lithium cobalt oxide; lithium ion; electrolyte; copper foil; aluminium foil; polyethylene separator; pvdf binder; carbon nanotube; nafion; activated carbon;
adhesive; sealant; polyurethane adhesive; epoxy adhesive; structural adhesive; butyl tape; plastisol; clear coat; e-coat; zinc phosphate; zinc nickel; chrome plating; electroless nickel; molybdenum disulfide; dlc; titanium nitride;
coolant; ethylene glycol; propylene glycol; engine oil; brake fluid; grease; gear oil; refrigerant}

Non-material lexicon for exclusion of those: \textit{material; materials; component; components; part; parts; housing; housings; casing; cover; covers; body; wiring; wire; wires; cable; cables; insulation; sealing; seal; seals; gasket; gaskets; connector; connectors; electronics; electronic; electrical; sensor; sensors; element; elements; coating; coatings; fastener; fasteners; other; various; additives; additive; lubricant; lubricants}

Mixtral terms: \textit{paint, vinyl, magnet}

GPT-4o terms: \textit{semiconductor, magnet}

\paragraph{Labeling.} A sentence is considered to mention the component if it contains at least two of the component's words, and at least one of them is not a generic word. A passage is direct evidence if one sentence within a window of 2 neighboring sentences hits the component and another or the same sentence contains a material term. We further label direct passages with a composition phrase (``made of'', ``consists of'', ...). 

\paragraph{Limitations.} The analysis is automatically only. We did no compare agreement with a human expert in this case. Further note that ``direct evidence'' is stricter than the earlier “helpful passage” analysis (22\%), which also counted passages that support indirect reasoning.

\begin{table*}[h]
\centering
\small
\begin{tabular}{l|rr|rr}
\toprule
              & \textsc{Cov}@1        & \textsc{Cov}@3        & \textsc{Prec}@1       & \textsc{Prec}@3       \\
\midrule
Baseline LLM  & \textbf{100.0\%} & \textbf{100.0\%} & 97.2\%           & \textbf{71.4\%}  \\
No RAG        & \textbf{100.0\%} & 88.0\%           & 99.0\%           & 63.9\%           \\
RAG           & 96.0\%           & 87.0\%           & \textbf{100.0\%} & 65.8\%           \\
CoVe          & \textbf{100.0\%} & 88.0\%           & 99.1\%           & 63.4\%           \\
\bottomrule
\end{tabular}
\caption{\textsc{coverage}@$n$ (Cov) and \textsc{precision}@$n$ (Prec) (Task 1).}
\label{tab:cov-prec-n}
\end{table*}

\begin{table*}[h]
\centering
\small
\begin{tabular}{l|r|rr|rr}
\toprule
 & \textbf{IAA} & \textbf{Cov@1} & \textbf{Cov@3} & \textbf{Prec@1} & \textbf{Prec@3} \\
\midrule
Batch 1 & 0.873 & 100.0\% & 96.6\%          & 98.9\%  & 79.8\% \\
Batch 2 & 0.450 & 100.0\% & 57.6\% & 98.1\%  & 15.6\% \\
Batch 3 & 0.891 & 100.0\% & 95.0\%          & 100.0\% & 83.7\% \\
Batch 4 & 0.703 & 100.0\% & 96.7\%          & 99.7\%  & 55.1\% \\
Batch 5 & 0.904 & 100.0\% & 98.3\%          & 100.0\% & 85.7\% \\
\bottomrule
\end{tabular}
\caption{Avg. IAA, \textsc{Coverage}@$n$ (Cov), and \textsc{Precision}@$n$ (Prec) per batch (Task 1).}
\label{tab:iaa-per-batch}
\end{table*}

\begin{table*}[h]
\centering
\small
\begin{tabular}{l|rrr|rr|r}
\toprule
                & No RAG  & RAG     & CoVe    & Mixtral & GPT-4o & Average \\
\midrule
First Material  & 88.0\%  & 88.5\%  & 90.0\%  & 86.7\%  & 91.1\%  & 88.9\% \\
Last Material   & 78.0\%  & 91.7\%  & 86.0\%  & 83.3\%  & 87.0\%  & 85.1\% \\
\bottomrule
\end{tabular}
\caption{\textsc{Precision}@2 for first and last material prediction of output (Task 1).}
\label{tab:domain-rag-first-last}
\end{table*}

\begin{table*}[t]
\centering
\small
\begin{tabular}{lccc}
\toprule
 & \textbf{Material rating} & \textbf{Explanation rating} & \textbf{Rank position} \\
 & ($\uparrow$ better) & ($\uparrow$ better) & ($\downarrow$ better) \\
\midrule
\multicolumn{4}{l}{\emph{Further contrasts (averaged over models)}} \\
CoVe $-$ RAG       & $+0.10$ [$-0.00$, $+0.21$] & \textbf{$+0.20$ [$+0.10$, $+0.30$]} & $-0.10$ [$-0.49$, $+0.29$] \\
GPT-4o $-$ Mixtral & $-0.08$ [$-0.17$, $+0.02$] & $+0.02$ [$-0.08$, $+0.12$] & $+0.12$ [$-0.18$, $+0.43$] \\
\midrule
\multicolumn{4}{l}{\emph{Per model}} \\
RAG $-$ No-RAG (Mixtral)  & $-0.13$ [$-0.27$, $+0.00$] & \textbf{$-0.27$ [$-0.43$, $-0.13$]} & $+0.29$ [$-0.17$, $+0.77$] \\
CoVe $-$ No-RAG (Mixtral) & $-0.11$ [$-0.24$, $+0.03$] & $-0.05$ [$-0.18$, $+0.07$] & $+0.16$ [$-0.35$, $+0.69$] \\
RAG $-$ No-RAG (GPT-4o)   & \textbf{$-0.35$ [$-0.55$, $-0.17$]} & \textbf{$-0.33$ [$-0.51$, $-0.16$]} & \textbf{$+0.77$ [$+0.21$, $+1.33$]} \\
CoVe $-$ No-RAG (GPT-4o)  & \textbf{$-0.18$ [$-0.31$, $-0.04$]} & \textbf{$-0.16$ [$-0.31$, $-0.01$]} & \textbf{$+0.70$ [$+0.19$, $+1.21$]} \\
\midrule
\multicolumn{4}{l}{\emph{Interaction: (RAG $-$ No-RAG) for GPT-4o minus the same for Mixtral}} \\
95\% CI & \textbf{$-0.22$ [$-0.43$, $-0.01$]} & $-0.06$ [$-0.26$, $+0.14$] & $+0.47$ [$-0.19$, $+1.15$] \\
98.3\% CI (Bonferroni, $K{=}3$) & $-0.22$ [$-0.48$, $+0.04$] & $-0.06$ [$-0.31$, $+0.18$] & $+0.47$ [$-0.33$, $+1.30$] \\
\bottomrule
\end{tabular}
\caption{Exploratory, unadjusted 95\% CIs (paired bootstrap over 50 components, annotator means). Rank positions range from 0 (best) to 5 (worst). The interaction rows test whether the RAG effect differs between the two models; the last row uses a Bonferroni-adjusted level for these three tests.}
\label{tab:exploratory}
\end{table*}

\begin{tcolorbox}[
    colback=blue!8!white,
    colframe=black,
    arc=4pt,
    boxrule=0.8pt,
    left=8pt, right=8pt, top=6pt, bottom=6pt, fontupper=\small
] \label{box:cove-prompt}
\textbf{Prompt:}
Your task is to create verification questions based on the below original question and the baseline response. The verification questions are meant for verifying the factual accuracy in the baseline response. Output should be numbered list of verification questions. \\
\\
Actual Question: ``In the automotive context, which materials is the component ``charge-air cooler water sprayer'' made out of''
Baseline Response:
``1. Water: The primary material used in a charge-air cooler water sprayer is water. It is used
to spray onto the charge-air cooler to lower the temperature of the incoming air, which
enhances engine efficiency by increasing air density.\\
2. Aluminum: While the sprayer itself may use water, the charge-air cooler often uses
aluminum due to its lightweight properties and good thermal conductivity, though aluminum
is near its upper technological limit at high temperatures and pressures.\\
The water is used in the greatest amount as it is continuously sprayed, while aluminum forms
the structural component of the cooler.''\\
\\
Final Verification Questions:\\
\\
\textbf{Generated Response:}\\
1. Is water a material used in the construction of the charge-air cooler water sprayer, or is it
only the medium being sprayed?\\
2. Is aluminum specifically used in the construction of the charge-air cooler water sprayer, or
is it used primarily in the charge-air cooler itself?\\
3. Does the baseline response accurately distinguish between the materials used in the sprayer
system and those used in related components like the charge-air cooler?\\
4. Are there any other materials commonly used in the construction of a charge-air cooler
water sprayer that are not mentioned in the baseline response?\\
\end{tcolorbox} 

\begin{figure*}
    \centering
    \includegraphics[width=0.99\linewidth]{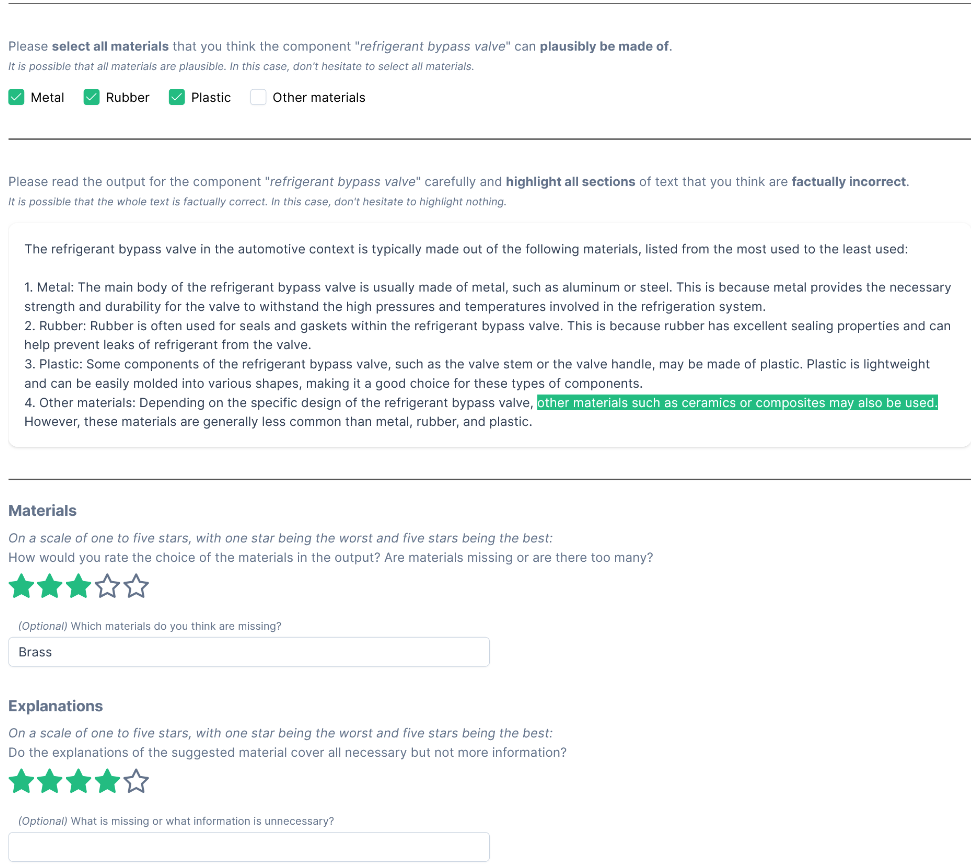}
    \caption{Screenshot of annotation interface for tasks 1--4. }
    \label{fig:wagner-ui-1}
\end{figure*}

\begin{figure*}
    \centering
    \includegraphics[width=0.99\linewidth]{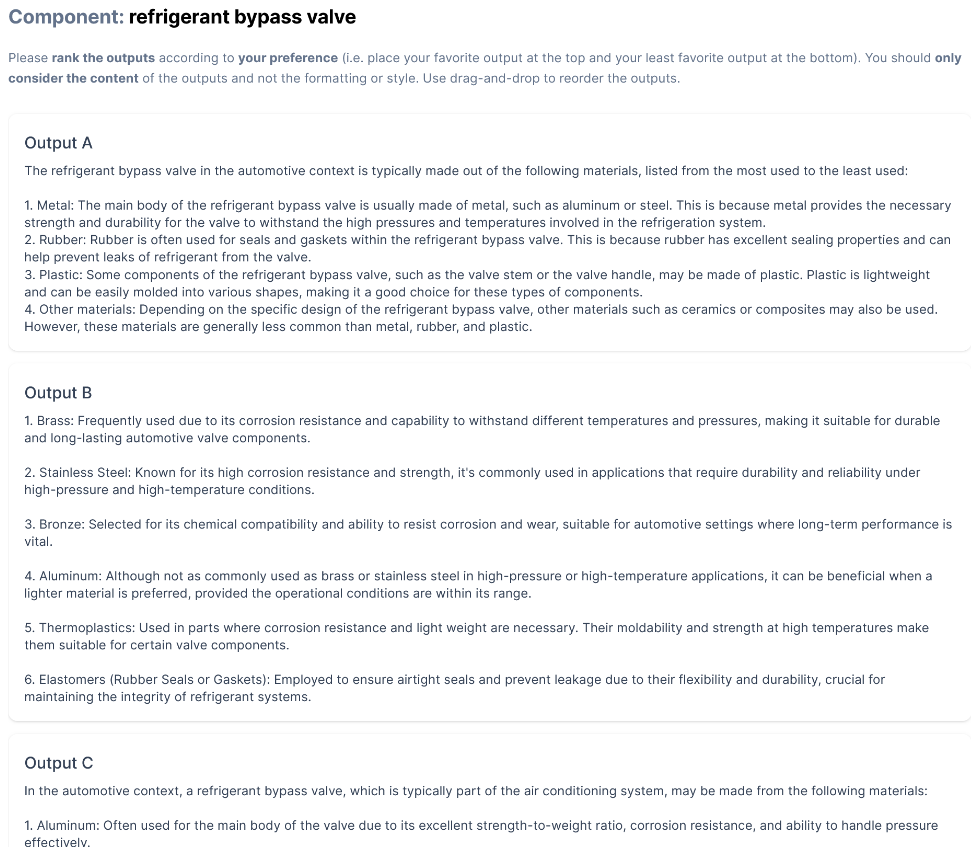}
    \caption{Screenshot of annotation interface for task 5. }
   \label{fig:wagner-ui-2}
\end{figure*}

\begin{table*}[t]
\centering
\small

\begin{tabular}{lrr}
\toprule
& \textbf{Mixtral} & \textbf{GPT-4o} \\
& Share (\%) [95\% CI] & Share (\%) [95\% CI] \\
\midrule
\multicolumn{3}{l}{\textit{Passage-level analysis}} \\
Direct evidence & 11.7 [9.1, 14.4] & 13.0 [10.1, 16.0] \\
Component only & 34.5 [30.4, 38.7] & 34.9 [30.7, 39.2] \\
Material only & 35.5 [32.6, 38.5] & 33.6 [30.6, 36.7] \\
Noise & 18.3 [14.8, 21.8] & 18.5 [15.1, 21.9] \\
\midrule
\multicolumn{3}{l}{\textit{Component-level analysis}} \\
Without any direct passage & 44\% & 43\% \\
With explicit composition statement & 17\% & 19\% \\
\bottomrule
\end{tabular}
\caption{Retrieval analysis for both models. Passage-level shares
are reported with 95\% confidence intervals (CI), bootstrap over components.
Component-level statistics are shown separately.}
\label{tab:retrieval-analysis}
\end{table*}

\end{document}